\documentclass[a4paper]{article}
\usepackage{epsfig}
\usepackage{citesort}
\usepackage{enumerate}
\usepackage{graphicx}  
\usepackage{color}     
\usepackage{color}
\usepackage{lscape}
\usepackage{amssymb}
\date{\today}
 \normalsize

\def\be {\begin{equation}}
\def\ee {\end{equation}}
\def\bea {\begin{eqnarray}}
\def\eea {\end{eqnarray}}
\def\bc {\begin{center}}
\def\ec {\end{center}}
\def\bfg {\begin{figure}}
\def\efg {\end{figure}}
\def\bi {\begin{itemize}}
\def\ei {\end{itemize}}
\def\nn {\nonumber}

\def\no {\noindent}

\def\vs {\vspace}

\def\a  {\alpha}
\def\b  {\beta}
\def\c  {\gamma}

\def\d  {\delta}

\def\l  {\lambda}

\def\m  {\mu}
\def\n  {\nu}
\def\o  {\omega}

\def\r  {\rho}
\def\th {\theta}
\def\s {\sigma}

\def\beq{\begin{equation}}
\def\eeq{\end{equatfion}}
\def\br{\begin{eqnarray}}
\def\er{\end{eqnarray}}
\newcommand{\eel}[1] {\label{#1}\end{equation}}

\newcommand{\bdm}{\begin{displaymath}}
\newcommand{\edm}{\end{displaymath}}
\begin{document}
\renewcommand{\thefootnote}{\fnsymbol{footnote}}

\vspace{.3cm}

\title{\Large\bf Comment on ``Quantum Raychaudhuri equation"}

\author
{ \it \bf  E. I. Lashin$^{1,2}$\thanks{slashin@zewailcity.edu.eg} and {\it \bf Djamel Dou${^3}$\thanks{djsdou@yahoo.com}}
\\
\small$^1$ Department of Physics, Faculty of Science, Ain Shams University,Cairo 11566,
Egypt.\\
\small$^2$ Centre for Fundamental Physics, Zewail City of Science and
Technology,\\
\small Sheikh Zayed, 6 October City, Giza 12588, Egypt.\\
\small$^3$ Department of Physics, College of Exact Sciences, El Oued University, El Oued 39000 , Algeria.}
\maketitle

\begin{center}
\small{\bf Abstract}\\[3mm]
\end{center}
We address the validity of the formalism and results presented in [ S. Das, Phys. Rev. {\bf D89} 084068 (2014)] with regard to quantum  Raychaudhuri equation.  The author obtained the so called quantum  Raychaudhuri equation  by replacing classical
geodesics with quantal trajectories arising from Bhommian  mechanics.  The resulting modified equation was used to draw some conclusions about the
inevitability   of focussing and the formation of conjugate points and  therefore singularity. We show  that the whole procedure is full of problematic points, on both physical relevancy and mathematical correctness. In particular, we illustrate the problems associated with the technical derivation of the so called quantum  Raychaudhuri equation, as well as its invalid physical implications.
\begin{minipage}[h]{14.0cm}
\end{minipage}
\vskip 0.3cm \hrule \vskip 0.5cm
\maketitle

The central finding in \cite{sd}  can be summarized as follows: Instead of  considering geodesics curves one  uses   modified curves
arising from Klein-Gordon-type equation of the following form (complex scalar field coupled non minimally to gravity),
 \be
\left(\Box  - {m^2 c^2\over \hbar} + \epsilon\, R \right) \Phi=0,\;\;\; \Box \equiv
g^{\m\n} D_\m D_\n.
\label{sc1}
\ee
The polar decomposition of the scalar field $\Phi$ is utilized as $ \Phi = {\cal R}\, \exp{\left(\displaystyle{{i S\over \hbar}}\right)}$ and then  using  Bhom's interpretation to define new congruence with  tangent vector field given by
\be
  u^\m ={d\,x^\m\over d\,\l}= {D^\m S \over m},
 \label{udef}
\ee
and consequently the norm of the velocity field $u^\m$ is given by
\be
u^\m u_\m = - c^2 +{\hbar^2\over m^2} {\Box {\cal R}\over {\cal R}} +{\hbar^2\over m^2} \,\epsilon\, R,
\label{nor}
\ee
form which the acceleration field follows as
\begin{equation}
\label{acc}
     a^\mu \equiv  u^\r\, D_\r u^\m   = {\hbar^2\over 2 m^2}\left\{ D_\m\!\!\left({\Box {\cal R}\over {\cal R}}\right) +\epsilon\, D_\m R\right\}.
\end{equation}
If we compare this resultant acceleration with the corresponding one given in \cite{sd}, we find a missing  factor ${1/ 2}$.

In deriving the   quantum version of Raychaudhuri equation (QRE) associated to the Bohmian trajectories, an essentially identical formalism devised for congruence generated by geodesics parameterized with the  proper time was  used    in \cite{sd}    to define the ``spatial" metric as $ h_{\m\n} = g_{\m\n} +  u_\m  u_\n $ and its corresponding projector $h^\m_{\;\;\n}$, the  expansion scalar
$ \th = h^{\m\n}D_\n\, u_\m$, the shear tensor $\s_{\m\n} =  \displaystyle{{1\over 2}}\,\left(D_\m\, u_\n  + D_\n\, u_\m\right)-\displaystyle{{1\over 3}}\, \th\, h_{\m\n} $, the rotation tensor  $ \o_{\m\n} =  \displaystyle{{1\over 2}}\,\left(D_\m\, u_\n  - D_\n\, u_\m\right) $,  and the following
equation which governs the evolution of the scalar expansion  was obtained:
\be
{d \th \over d\l } +{1\over 3}\, \th^2  + \s_{\m\n}\, \s^{\m\n}  - 2\,h^{\m\n}\, D_\m\, a_\n  + R_{\m\n}\, u^\m \, u^\n=0.
\label{raysd2}
\ee
Before discussing the correctness of this QRE presented in Eq. (\ref{raysd2}), let us mention few important remarks and  arguable points about the whole approach which can be summarized as follows:
\begin{enumerate}[(i)]
     \item The norm of the tangent field $u^\m$,  as found in Eq.(\ref{nor}),  does not have  a  well defined sign. The expected deviation from $\left(-c^2\right)$ is considered to be small, but there seems to be no control on the magnitude of this deviation, especially when the curvature becomes large, let alone the unknown sign of the term $\displaystyle {\hbar^2\over m^2} {\Box {\cal R}\over {\cal R}}$.
  Therefore, the points along the trajectory can not be  chronologically ordered in a unique way which is independent of any reference frame. This violation of causality can not be reconciled in any reasonable manner. Although nonrelativistic quantum mechanics allows for causality violation, while relativistic theory softens this violation, the quantum field theory completely evades the causality violation in a miraculous and well-known way \cite{peskin,bank} through the exact cancelation between the contribution of particles and that of antiparticles. This serious flaw was pointed out  in \cite{lash} concerning the cosmological study
based on  QRE carried out in \cite{fd}.

\item The projector $h_{\m\n}$  is not properly defined in terms of normalized tangent vector $\hat{u}^\m$; more precisely it cannot be taken as a spatial metric nor $h^\m_{\;\;\n}$ is a projector onto the subspace of tangent space perpendicular to $ u^\m$ .

 \item The resulting tensor $D_\n u_\m$ is no longer purely spatial unless the congruence generated by $u^\m$ is a timelike geodesic.

\item  The expansion scalar $\th$   will no longer have the same interpretation as,  for instance, a measure of the fractional rate at which the volume of a ball of matter changes with respect to time as measured by a central comoving observer.

\item The tensor $\s_{\m\n}$  is not any longer purely spatial and consequently  $ \s_{\m\n}\, \s^{\m\n}$ will
 not have a fixed signature; therefore,  the positivity energy conditions cannot be used without considering  the signature of this term.
 \end{enumerate}
It thus results  that the whole approach is ill defined and loses a considerable part of its geometrical and physical meaning.

As to the  correctness  of the modified Raychaudhuri equation (QRE) derived in \cite{sd}, i.e., Eq.~(\ref{raysd2}), and
apart from the factor $2$,  which is related to the factor $\displaystyle{{1/2}}$ mentioned before, it seems that the author of \cite{sd}  took only into account the modification
 arising from   $D_\m a^\m$, which is non vanishing by virtue of Eq.~(\ref{acc}). However it should be mentioned  that
not all the results obtained for geodesic congruence nor the formalism carry over directly to the case of congruence generated by timelike curves. Indeed,
there are many other terms  coming from the facts that $D_\r h_{\m\n}\neq 0$, $h^\m_{\;\;\n}$ is no longer projector nor $h_{\m\n}$ has trace equals to $3$, and $\sigma_{\m\n}h^{\m\n}\neq 0$.  Now, even though we find the whole formalism used in \cite{sd} ill defined,  it would be interesting  for the sake of completeness  that we give the correct ``formal" equation that would arise  by taking into accounts all the terms. The resulting modified Raychaudhuri equation turns out to be
\be
{d \th \over d\l }- a_\n\, a^\n- \displaystyle{{1\over 9}}\, \th^2\, h_{\m\n}\, h^{\m\n}+\displaystyle{{2\over 3}}\,\th^2  + \s_{\m\n}\, \s^{\m\n}  - h^{\m\n }\, D_\m\, a_\n  + R_{\m\n}\, u^\m\, u^\n=0.
\label{raysd1}
\ee
Comparing this finding in Eq. (\ref{raysd1}) with that of  Eq. (\ref{raysd2}), we find  that the two equations are completely different with the exception of the last three terms.

Although the mathematical approach devised in \cite{sd} leads to incorrect modified Raychaudhuri equation, there is still a correct modified equation which can be deduced for the Bohmian approach proposed therein; of course we leave aside the problems with Bohm's interpretation and its relativistic generalization.

It turns out that there are two equivalent formalisms which can be used to write a modified Raychaudhuri equation for the congruence generated by the velocity field satisfying Eqs. (\ref{udef}) and (\ref{nor}). The first approach is to use directly the non-normalized velocity field (assuming that it remains timelike) as given by Eqs.(\ref{udef}) and (\ref{nor}) and define all mathematical quantities properly. This approach is slightly different from the standard approach  and thus one has to derive the corresponding Raychaudhuri equation from  scratch. The proper spatial metric can be defined as
   \begin{equation}\label{spm}
   h_{\m\n}= g_{\m\n} -\frac{u_\m u_\n}{u_\a u^\a}.
   \end{equation}
The expansion tensor   is defined by projecting the symmetric part of $ D_\m u_\n $   onto the orthogonal space (in our case $ D_\m u_\n $ is already symmetric),
\begin{equation}\label{dec1}
   \Theta_{\m\n}= h_\m^{~\a}h_\n^{~\r}D_\a u_\r.
   \end{equation}
The resulting expansion tensor can be seen to be purely spatial, i.e., $ \Theta_{\m\n}\, u^\m=\Theta_{\m\n}\, u^\n= 0$, and
the expansion scalar can then be defined as $\Theta= h^{\m\n}\Theta_{\m\n}=g^{\m\n}\Theta_{\m\n}$. The expansion tensor $\Theta_{\m\n}$  can be decomposed in terms of its irreducible parts as
\bea
\label{dec2}
   \Theta_{\m\n } = \sigma_{\m\n} + \frac{\Theta}{3} h_{\m\n}, && \mbox{where},\;\;\sigma_{\m\n}= \Theta_{\m\n }-\frac{\Theta}{3} h_{\m\n}.
\eea
It is straightforward to show that the scalar expansion satisfies the following modified Raychaudhuri  equation ( or QRE):
\be
{d {\Theta} \over d \l }+\displaystyle{{1\over 3}}\, \Theta^2  + {\s}_{\m\n}\, {\s}^{\m\n}  - {h}^{\m\n }\, D_\m a_\n +3\,h^{\m\n}\,\displaystyle{\frac{a_\m\, a_\n}{u^2}} + R_{\m\n} \,u^\m\, u^\n=0.
\label{mre1}
\ee
Except for the term $h^{\m\n } D_\m a_\n$ (up to a factor 2) our equation is  substantially different from the equation obtained in \cite{sd}, i.e., Eq. (\ref{raysd2}).
Moreover the tenor $\sigma_{\m\n}$ and the projector $h_\m^{\;\;\n}$ are very different from the ones given in \cite{sd}. Actually  it is the fact that $\sigma_{\m\n}$ is being purely spatial  which makes  the positive energy conditions of particular importance.

The second alternative approach to derive a modified Raychaudhuri equation for the velocity filed $u^\m$ is to normalize it viz $\displaystyle \hat{u}_\m = {D_\m S \over \sqrt{ \left|D_\a S D^\a S\right|}}$. Having a normalized timelike velocity field $\hat{u}_\m$ together with all necessary derived quantities such as spacial projector $h_{\m\n}$,
$\th$ (expansion), $\s_{\n\m}$ (shear tensor), $\o_{\n\m}$ (rotation tensor) and $a^\mu$ (acceleration) which are defined as
\be
\left.
\begin{array}{lll}
h_{\m\n} &=& g_{\m\n} +  \hat{u}_\m\,  \hat{u}_\n,\;\;\mbox{where}\;\; \hat{u}_\a  \hat{u}^\a = -1,\\\\
\th &=& D_\m\, \hat{u}^\m,\\\\
\s_{\n\m} &= & \displaystyle{{1\over 2}}\,h^\r_{\;\n}\, h^\s_{\;\m}\, \left(D_\r\, \hat{u}_\s  + D_\s\, \hat{u}_\r\right) -\displaystyle{{1\over 3}}\, \th\, h_{\m\n},
\\\\
\o_{\n\m} &= & \displaystyle{{1\over 2}}\,h^\r_{\;\n}\, h^\s_{\;\m}\, \left(D_\r\, \hat{u}_\s  - D_\s\, \hat{u}_\r\right),\\\\
a^\mu &=& \hat{u}^\r\, D_\r \hat{u}^\m.
\end{array}
\right\}
\label{def1}
\ee
It is straightforward to use the standard formalism devised for a generic timelike curve  as described in \cite{haw,senov1}  and derive the following corrected or modified Raychaudhuri equation that governs the evolution of the expansion parameter $\th$ as
\be
{d \th \over d\l }+\displaystyle{{1\over 3}}\, \th^2  + \s_{\m\n}\, \s^{\m\n}  - \o_{\m\n}\, \o^{\m\n}- D_\m\, a^\m  + R_{\m\n}\, \hat{u}^\m\, \hat{u}^\n=0.
\label{mre2}
\ee
This equation is again completely different from  Eq. (\ref{raysd2} ). As can be seen, the rotation tensor is no longer vanishing and the other quantities differs in their definition from the ones given in \cite{sd}.

Another incidence which we see presenting a problem in the approach of QRE is the one related to Jacobi equation.  By using the standard definition for the  relative acceleration between geodesics and applying it to the case of nongeodesic timelike curves, where the term $ D_\b \left(u^\c D_\c u^\m\right)$ is not vanishing, one gets the modified Jacobi equation,
\bea
{D^2 \eta^\m \over d \l^2} &\equiv& u^\c\,D_\c\left(u^\b D_\b\eta^\m\right),\nn\\
&=& \eta^\b D_\b \left(u^\c D_\c u^\m\right)- R^\m_{\;\:\a \b \c}\, u^\a u^\c \eta^\b,\nn\\
 &=&- R^\m_{\;\:\a \b \c}\, u^\a u^\c \eta^\b + {\hbar^2\over 2  m^2}\,\eta^\b D_\b D^\m \left\{ \left({\Box {\cal R}\over {\cal R}}\right) +\epsilon\, R\right\}.
 \label{jsd1}
\eea
Again, this is to be compared to the different result in \cite{sd}( after setting $\epsilon=0$) :
\be
{D^2 \eta^\m \over d \l^2} = - R^\m_{\;\:\a \b \c}\, u^\a u^\c \eta^\b +{\hbar^2\over  m^2}\,\eta^\c\,D_\c \, D^\m \left({\Box {\cal R}\over {\cal R}}\right).
\label{jsd2}
\ee
We  again note that  both equations, Eq. (\ref{jsd1}) and  Eq. (\ref{jsd2}),  do not correspond correctly to the deviation of timelike curves  for reasons similar to the ones we mentioned for the  Raychaudhuri equation. The author of \cite{sd} used the same definition for the  relative acceleration between geodesics and took only into consideration the nonvanishing of $  D_\b \left(u^\c D_\c u^\m\right) $; however,   to properly derive  the Jacobi equation for timelike curves,  the deviation vector $\eta^\m$ has to be defined in a more subtle way by considering the relative separation
$\eta^\m$ modulo a component  parallel to $\hat{u}^\n$  and thus the projection of $\eta^\m$  i.e  $ \eta^\m_\bot = h^\m_{\;\;\n} \eta^\n $. This subtlety does not arise in case of   geodesic
congruence because $\eta^\m$  and $\hat{u}^\n$ can always be chosen to be orthogonal along the geodesic. For affinely parameterized  timelike curves with parameter $\l$, the deviation equation, after lengthy algebraic manipulations,  is given by \cite{haw}
\bea
    h^\m_{\;\;\a} \frac{D}{d\l} ( h^\a_{\;\;\b}\frac{D}{d\l}\eta^\b_\bot)&\equiv &
     h^\m_{\;\;\a} \hat{u}^\d D_\d \left[ h^\a_{\;\;\b}\left(\hat{u}^\c D_\c \eta^\b_\bot\right)\right],\nn\\
   &=& -R^\m_{\;\;\a\b\c} \eta^\b_\bot  \hat{u}^\a \hat{u}^\c
    + h^\m_{\;\;\a} \left( D_\b a^\a\right) \eta^\b_\bot + a^\m a_\a \eta^\a_\bot,
\label{corrjac}
\eea
where $a_\m$ is the acceleration as defined in Eq. (\ref{def1}).
It is obvious that if one considers timelike curves with non normalized tangent vector $\left( u^\a u_\a \neq -1\right)$ then the deviation Eq. (\ref{corrjac}) would have extra terms on the right hand side. Clearly, using non-normalized velocity fields introduces unnecessary complications without any benefit.

Let us finally comment on the  physical conclusions drawn by the author  of \cite{sd}, leaving aside the problems with Bohm's interpretation, the correctness of the  modified Raychaudhuri equation  and  the fact that the formalism itself is ill defined. The  claim was  that the modification brought to Raychaudhuri
equation naturally prevents focussing and the formation of conjugate points. The main argument was the presence of a quantum potential and the fact that congruence  was determined through first order differential equation $\displaystyle \left({d\,x^\m\over d\,\l}={D^\m S \over m}\right)$ and so the uniqueness would prevent caustic points to exist.

It is true that quantal trajectories available for a single particles in nonrelativistic  Bohmian mechanics form a congruence due to the fact that their velocity field is given by ${\mathbf \nabla} S$ (which is a single-valued function), but this by no means ensures that they will preserve this property (maintain  congruence) in the presence of gravity. Indeed this what Raychaudhuri equation and singularity theorem are  all about. A congruence in an open region of spacetime  is by definition a family of curves such that through each point in this region there passes  one and only one curve from this family. The singularity theorem that is generally proven, as presented in \cite{pen2}, starts with a congruence, in particular with a rotationless congruence (or hypersurface orthogonal) and therefore with a velocity field given by $ u^\mu = D^\mu f$ ( Frobenius theorem), and under some energy positivity conditions  it is shown that this family of curves will after a finite proper time fail to be congruence and  geodesics get focused as a result of gravity attraction. This is shown by studying Raychaudhuri equation which govern the development of the congruence.

Therefore in order to conclude that focusing never happen within a finite proper time one must follow the development of the velocity field using the new
Raychaudhuri equation, by analyzing the different competing terms appearing in the equation before any conclusion can be drawn. This  has never been done by the author. Similar remarks applies for Jacobi equation.

Going further now and assuming that there are actually no caustic points for those Bohmian trajectories, this would not change the situation for timelike or null geodesic completeness of the given spacetime. The reason that timelike or null geodesics are important is that  the first ones may correspond to a physical observer, while the second ones correspond to a ray of light. In fact, one can extend the notion of completeness to a general timelike curve which can also represent a physical observer, and this extension turns out to be necessary since there are spaces  which are geodesically complete but incomplete for timelike curves \cite{ger}.

On the other hand, it is intriguing to note that  the equations deduced for Bohmian trajectories are just for  test particles
  moving under the action of presumed quantum force while keeping  the  underlying spacetime geometry untouched. This alone can't evade singularity since,  for instance, behind black hole horizon neither physical matter nor physical light can escape singularity. Now, even if the  conclusions  of the author are accepted as they stand,  it must be mentioned that such quantum effects should be expected to affect the convergence of timelike geodesics only at very small distances. In such regions the  curvature  becomes so extreme that it might well count as a singularity;  however, it is generally believed  that general relativity itself will break down before the final singularity is reached.

To sum up, the QRE approach presented in \cite{sd} does not evade singularity, and  the whole formulation is, in our opinion, troublesome on both mathematical and physical sides.
This certainly invalidates the whole subsequent works \cite{fd,sdc1,sdc2,sdc3,fk,ar} that adopted the QRE approach and built on it.
We believe that this QRE question needs to be fully addressed and settled before any further  works based on QRE are contemplated. We hope that the present work will draw attention to this issue, which, somehow, is important in the arena of quantum general relativity.

\vs{.2cm}
\no {\bf Acknowledgment}
\no
We thank N. Chamoun for useful discussions.
%



\end{document}